# Penn & Slavery Project's Augmented Reality Tour

Augmenting a Campus to Reveal a Hidden History

VanJessica J Gladney

PhD Candidate, University of Pennsylvania, vgladney@sas.upenn.edu

Breanna N Moore

PhD Candidate, University of Pennsylvania, moorebre@sas.upenn.edu

Kathleen M Brown

David Boies Professor of History, University of Pennsylvania, kabrown@sas.upenn.edu

In 2006 and 2016, the University of Pennsylvania denied any ties to slavery. In 2017, a group of undergraduate researchers, led by Professor Kathleen Brown, investigated this claim. Initial research, focused on 18th-century faculty and trustees who owned slaves, revealed deep connections between the university's history and the institution of slavery. These findings, and discussions amongst the researchers shaped the Penn & Slavery Project's (PSP) goal to redefining "complicity" beyond ownership. Breanna Moore's contributions in PSP's second semester expanded the project's focus to include generational wealth gaps. In 2018, VanJessica Gladney served as the PSP's Public History Fellow and spread the project outreach in the greater Philadelphia area. That year, the PSP team began to design an augmented reality app as a "Digital Interruption" and an attempt to display the truth about Penn's history on its campus. Unfortunately, PSP faced delays due to COVID-19. Despite setbacks, the project persisted, engaging with activists and the wider community to confront historical injustices and modern inequalities.

CCS CONCEPTS • Human-Centered Computing~ *Human computer interaction (HCI)* ~ Interaction paradigms ~ Mixed/Augmented Reality • Applied Computing ~ Arts and Humanities • Applied Computing ~ Education ~ Digital Libraries and Archives

**Additional Keywords and Phrases:** Augmented Reality, Digital Interruption, Repairing Historical Erasure, Reparative Justice

# 1 INTRODUCTION

In 2006 and 2016 the University of Pennsylvania denied having any connections to the institution of slavery. [1, 2] In 2017, a group of undergraduate researchers, under the guidance of Professor Kathleen Brown, turned to the Penn Archives to research that claim. During the first semester of the independent study, their research focused on the 18th-century faculty and trustees who owned enslaved people. In their first semester, they discovered that at least 75 18th-century trustees and faculty owned enslaved people and profited from the labor of one enslaved man on Penn's original campus. [3] Over that semester, the students identified threads connecting the university to the institution of slavery in ways that expanded past ownership. As the Penn & Slavery Project began to take shape, the researchers made it a project goal to complicate the definition of '*complicity*' from one that focuses on ownership, to one that "speaks to the many ways in which colonial universities relied on and contributed to America's slave society in the years prior to the Civil War." [4] They knew their work was not complete, and so they kept researching. [5]

In their second semester, they branched out from researching ownership to focus on how Penn shaped and was shaped by slavery in its early financial practices, the medical education students received in the school's earliest days, and the presence of the legacy of slavery on its original and current campus.[6] That same semester, the project was further strengthened when Breanna Moore, a Penn Alum joined the project. She had spent the past few years conducting independent research on the Penn family who enslaved her ancestors and producing a documentary to tell her family's story. She added her voice to the project, emphasizing the ways educational opportunities served as an indicator of the gap in generational wealth created by the institution of slavery. Her contributions encouraged the members to think of the Penn & Slavery Project as more than a series of research papers, but a method to interrogate the ways the institution of slavery has created the injustices and inequality we see in the modern-day United States of America.

In 2018, the Penn & Slavery Project focused on making their work as public-facing as possible. VanJessica Gladney received a Public History Fellowship from the university's Provost's Office. During her fellowship, she gave presentations about the project throughout the greater Philadelphia area and built the project's website to feature student research. [7] That same year, the Penn & Slavery Project began designing an augmented reality mobile application to help project the truth about Penn's history onto its campus. The mobile application's roll-out was scheduled for 2021, but due to COVID-19 restrictions, the event had to be converted to a virtual presentation, which included this informative video, and a digital exhibit about the tour. [8,9]

## 2 THE AUGMENTED REALITY TOUR

The Penn & Slavery Project's Augmented Reality tour offers a groundbreaking exploration into the intricate connections between Penn and slavery. Comprising six immersive stops, each featuring meticulously researched exhibits crafted by undergraduate scholars, the tour delves deep into the university's historical entanglements with the institution of slavery. The stops were chosen deliberately to reveal a specific part of the university's history as it relates to the location on campus. The intention was to perform a 'digital interruption' of the story the university tells about itself by using historically informed Augmented Reality Exhibits to add context about its connections to slavery to the existing iconography on campus. Through the innovative use of Augmented Reality technology, visitors are transported into a digital realm where they confront the stark realities of Penn's past. From uncovering the stories of enslaved individuals to examining the complicity of academic figures in perpetuating racial hierarchies, each exhibit sheds light on a different facet of Penn's complex history. As participants journey through the tour, they are challenged to confront uncomfortable truths and engage in critical dialogue about the enduring legacies of slavery within academia and society at large.

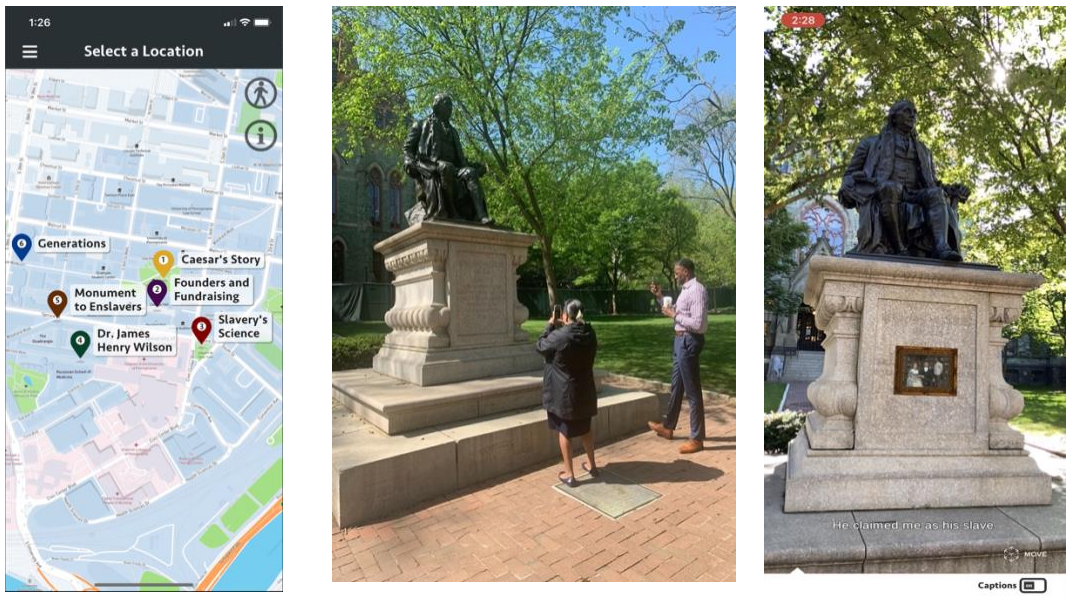

Figure 1: The first stop on the tour, Caesar's Story, projects a reimagined historical portrait onto a statue of Benjamin Franklin that sits at the heart of the campus of the University of Pennsylvania. When users look at the statue while using the Penn & Slavery Project Augmented Reality Tour Mobile Application, they will see the portrait come to life and hear a reenactor tell Caesar's story.

### 2.1 Artifacts

The Penn & Slavery Project Augmented Reality app was conceived as a groundbreaking tool to disrupt the traditional narrative of the university campus, serving as a "digital interruption" that unveils the hidden legacy of slavery. Through the innovative use of Augmented Reality technology, the app aimed to peel back the layers of history, exposing the often-overlooked contributions and experiences of Black individuals within the university's past. By superimposing digital elements onto physical spaces, the app sought to create immersive experiences that brought these forgotten voices to the forefront, challenging prevailing notions of historical erasure and exclusion. Through its dynamic and interactive features, the app not only shed light on the pervasive impact of slavery but also sparked crucial conversations about representation, memory, and accountability within the university community and beyond. The following are some of the notable artifacts included in the tour.

**Helping Hands|** featured in the [Dr. James Henry Wilson](#) stop.

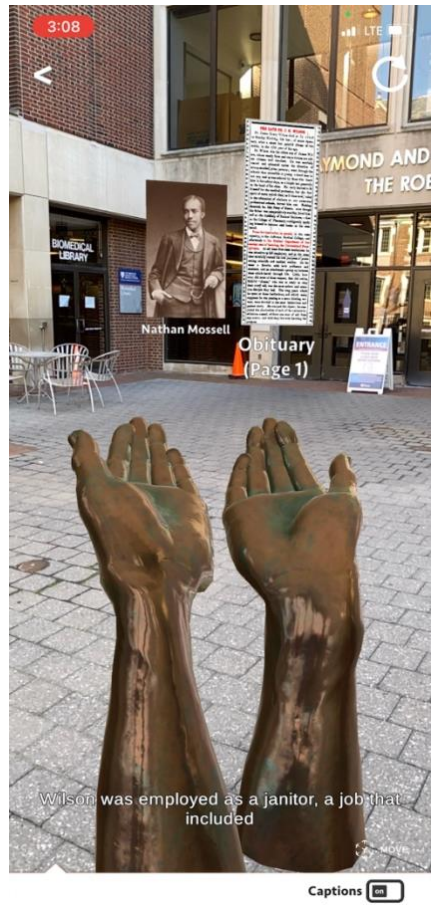

Figure 2: With the Penn & Slavery Project App, users will. see this Augmented Realty statue commemorating the medical achievements of Dr. James Henry Wilson, which have yet to be recognized by the university. Users can also interact with the archival documents floating above the statue of Dr. Wilson's "Helping Hands."

The Penn & Slavery Project not only aims to rectify the historical record regarding Penn's ties to slavery but also prioritizes the elevation of African American narratives and the amplification of Black voices. At the stop featuring Dr. James Henry Wilson, situated at the Robert Wood Johnson Pavilion (the home of Penn's Biomedical library, classrooms, and offices) and researched by Bryan Anderson-Wooten and Dallas Taylor, this dual objective is effectively realized. Dr. Wilson, an African American individual who pursued medical studies at Penn during the 1840s, faced discrimination and was ultimately denied a degree. Despite this setback, he established a successful medical practice and collaborated with esteemed physicians in Philadelphia. Remarkably, the Penn Archive had inaccurately merged Dr. Wilson's records with those of another Black individual named Albert Wilson, an error rectified by diligent undergraduate researchers within the project.

The Penn & Slavery Project embarked on a groundbreaking endeavor to combat historical erasure and honor the contributions of Dr. James Henry Wilson through the creation of an Augmented Reality (AR) exhibit showcasing an imagined statue titled "Wilson's Helping Hands." This innovative exhibit serves as a powerful tribute to Dr. Wilson's pioneering work and enduring legacy within the Penn community. Through AR technology, users are transported into a virtual realm where they encounter a lifelike representation of the statue, depicting Dr. Wilson extending a helping hand to those in need. Above his hands, are the archival documents that reveal the truth about his unorthodox medical education at the University of Pennsylvania. The exhibit not only celebrates Dr. Wilson's remarkable achievements but also highlights his resilience in the face of adversity, including the discrimination he faced while pursuing medical studies at Penn in the 1840s. By immortalizing Dr. Wilson's compassionate spirit and unwavering dedication to serving others, "Wilson's Helping Hands" stands as a beacon of hope and a testament to the importance of acknowledging and commemorating marginalized voices in history.

**The Dome**| featured in the [Slavery's Science](#) stop.

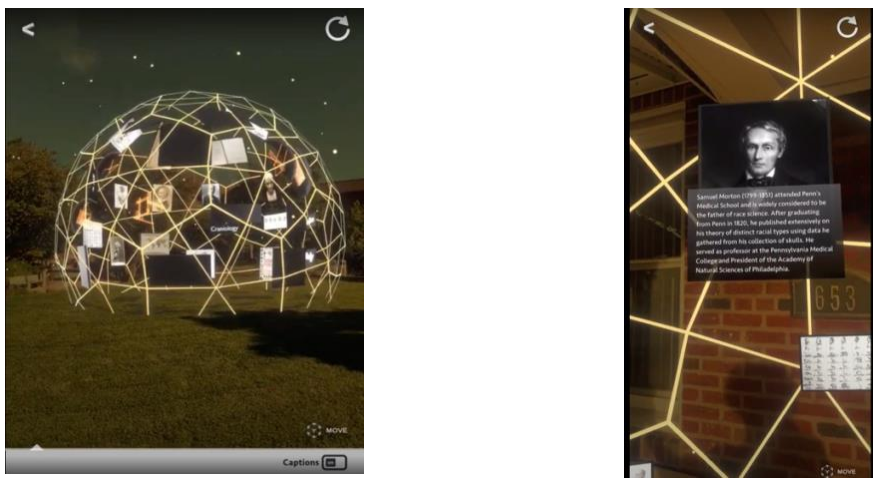

Figure 3: These are two screen capture of the dome that spawns in the Slavery Science stop, the images, text, and 3D-modeled artifacts inside the dome reveal the history of scientific racism prevalent in Penn's Early Medical school. In addition, here is a [link to a video of the dome spawning](#).

The Slavery's Science exhibit, researched by Carson Eckhard, Archana Upadhyay, and Paul Wolff Mitchell, offers public access to the history of Penn's Samuel Morton Skull Collection. Their diligent research prompted the Penn Museum to relocate the collection. Through the Augmented Reality (AR) exhibit, users are enveloped within a dome showcasing interactive artifacts, images, and terminology, elucidating the intertwined relationship between Penn's medical school professors and the propagation of scientific racism. This immersive experience underscores the pervasive nature of this history, emphasizing that the legacy of slavery permeates all aspects of society. Furthermore, the dome serves as a poignant reminder of the profound and enduring impact of Penn's historical complicity with slavery, particularly through the dissemination of medical and scientific racism. Figures such as Charles Caldwell, Samuel George Morton, and Benjamin Rush, who served as professors and instructors at Penn's nineteenth-century medical school, actively perpetuated notions of racial superiority through their teachings. Morton, in particular, utilized his position to espouse theories of polygenesis and craniometry, falsely correlating skull size with mental inferiority among Black individuals, thus reinforcing racial hierarchies. Following Morton's passing, his widow sold the collection to the Academy of Natural Sciences, which subsequently transferred it to the Penn Museum in 1966.

**The Moore Family Quilt**| featured in the [Generations](#) stop.

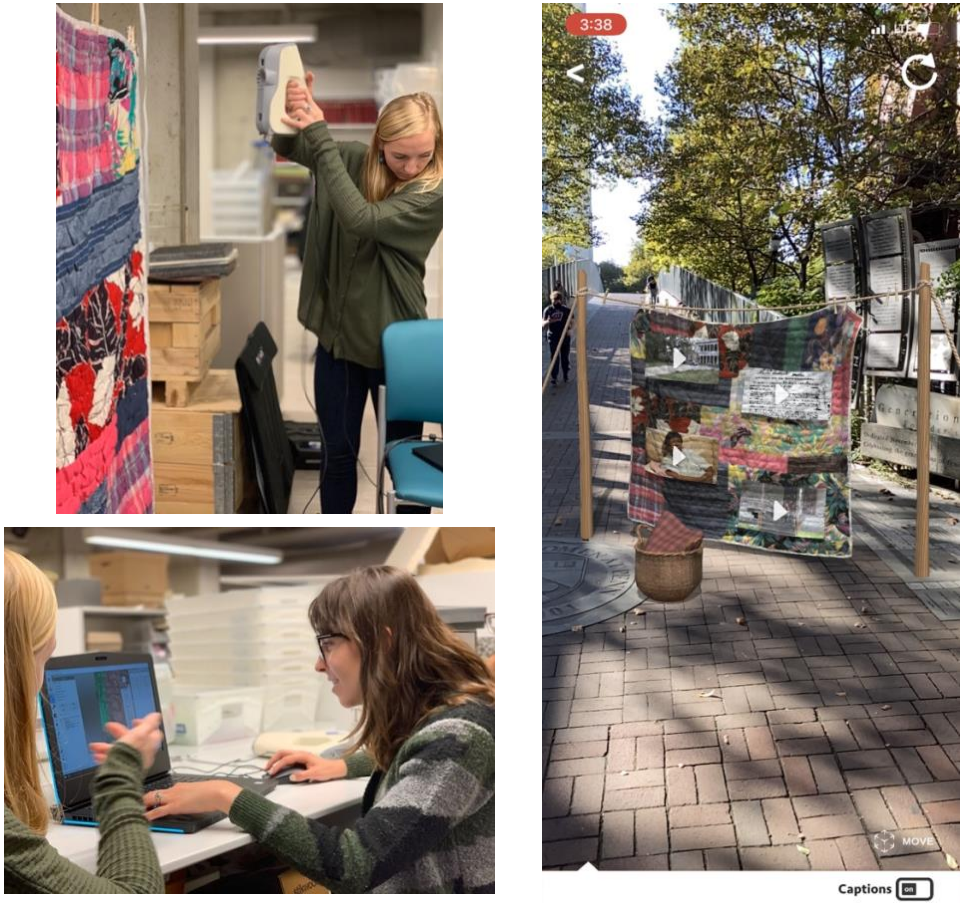

Figure 4: The Generations stop beautifully illustrates the illustrates the wide array of possibilities when using Augmented Reality as an educational tool. This was an amazing opportunity for interdisciplinary collaboration, blending materials history and graphic design. In addition, this is a multimedia stop, where, once modeled, the Penn & Slavery Project designers placed clips of the documentary Breanna Moore produced onto the squares of the quilt, allowing her to tell her family's story.

Breanna Moore generously allowed the Penn & Slavery Project to scan, and 3D Model her family quilt, for the focus of the final stop of the tour, which features the research she conducted on the connections between Penn, slavery, and her family. The exhibit places her family's quilt in front of the Generations Bridge on Penn's campus, which has bricks engraved with the names of donors and alumni. This stop raises questions about whose legacies are remembered and deemed worthy of remembrance. It illustrates how academia and education are forms of generational wealth, and that slavery and white supremacy created a racialized gap in that wealth.

## 3 OUTCOMES

The outcomes of the Penn & Slavery Project have reverberated throughout the University of Pennsylvania and beyond, leaving an indelible mark on both academia and societal consciousness. The first of these was a symposium held on campus, a momentous gathering where scholars, students, and community members convened to delve into the project's findings. [10] This symposium not only provided a platform for discourse on the project's revelations but also served as a catalyst for further research initiatives exploring the intersection of history, race, and power dynamics. A pivotal result of student research within the project was the removal of the George Whitefield Statue, a symbolic gesture emblematic of the institution's recognition of the project's hard work and an official acknowledgement of its historical complicity in slavery. [11] This act of institutional introspection underscored the transformative potential of academic inquiry in driving tangible change.

Moreover, the project's impact transcended the confines of the campus, extending into the realm of activism through collaboration with local Philadelphia advocates. Together, project members and community activists mobilized to confront the troubling legacy embodied by the Morton Skull Collection housed within the Penn Museum. By amplifying marginalized voices and advocating for the collection's removal, they challenged entrenched power structures and propelled the discourse surrounding ethical museum practices and cultural restitution. The project's efforts did not go unnoticed, as it garnered widespread attention from the local press, thrusting issues of historical reckoning and social justice into the spotlight. [12, 13, 14, 15] This pressure led the university to apologize for their possession of the collection, and begin efforts towards repatriation and a respectful burial of the remains. [16, 17]

In addition to these impactful initiatives, the Penn & Slavery Project embarked on a journey of public engagement and education, offering guided tours on campus as a means of confronting the university's complex historical legacy. The inaugural tour, held on Juneteenth 2023, marked a poignant moment of reflection and commemoration, underscoring the project's commitment to fostering dialogue and promoting historical awareness. [18] Furthermore, the university's decision to join the "Universities Studying Slavery" organization signaled a broader commitment to collective action and collaboration within the higher education community. [19] Through these collective endeavors, the Penn & Slavery Project continues to catalyze meaningful change, challenging entrenched narratives and shaping a more inclusive and equitable future.